\documentclass[preprint,superscriptaddress, nofootinbib]{revtex4-1}
\usepackage{graphicx}
\usepackage{amsmath}
\usepackage{amssymb}
\usepackage[colorlinks,linkcolor=blue,
            citecolor=blue,
            hyperindex,
            pdfstartview=FitH,
            plainpages=false]
            {hyperref}
\usepackage{lineno}
\usepackage{color}
\usepackage{here}

\begin{document}

\title{Observation of a flat and extended surface state in a topological semimetal}

\author{Ryo~Mori}
\affiliation{Applied Science \& Technology, University of California, Berkeley, California 94720, USA}
\affiliation{Materials Sciences Division, Lawrence Berkeley National Laboratory, Berkeley, California 94720, USA}
\author{Kefeng~Wang}
\affiliation{Maryland Quantum Materials Center, Department of Physics, University of Maryland, College Park, Maryland 20742, USA.}
\author{Takahiro~Morimoto}
\affiliation{Department of Applied Physics, The University of Tokyo, Hongo, Tokyo, 113-8656, Japan}
\affiliation{JST, PRESTO, Kawaguchi, Saitama, 332-0012, Japan}
\author{Jonathan~D.~Denlinger}
\affiliation{Advanced Light Source, Lawrence Berkeley National Laboratory, Berkeley, California 94720, USA}
\author{Johnpierre~Paglione}
\affiliation{Maryland Quantum Materials Center, Department of Physics, University of Maryland, College Park, Maryland 20742, USA.}
\author{Alessandra~Lanzara}
\email[To whom correspondence should be addressed. Email: ]{ALanzara@lbl.gov}
\affiliation{Materials Sciences Division, Lawrence Berkeley National Laboratory, Berkeley, California 94720, USA}
\affiliation{Department of Physics, University of California, Berkeley, California 94720, USA}

\begin{abstract}
A topological flatband, also known as drumhead states, is an ideal platform to drive new exotic topological quantum phases. 
Using angle-resolved photoemission spectroscopy experiments, we reveal the emergence of a highly localized possible drumhead surface state in a topological semimetal BaAl$_4$ and provide its full energy and momentum space topology.
We find that the observed surface state is highly localized in momentum, inside a square-shaped bulk Dirac nodal loop, and in energy, leading to a flat band and a peak in the density of state.
These results establish this class of materials as a possible experimental realization of drumhead surface states and provide an important reference for future studies of fundamental physics of topological quantum phase transition.
\end{abstract}

\maketitle

One notable class of the topological non-trivial states---nodal line semimetals---are the neighbor states to various topological quantum phases, such as three-dimensional Dirac semimetals, Weyl semimetals, topological insulators, spinful Weyl nodal line semimetals, and hence it is regarded as an ideal platform to study and control the quantum topological phase transition by breaking symmetries\cite{Yang2018}.
In a topological nodal line semimetal, the linearly degenerate bands cross each other on a mirror plane giving rise to a nodal line in the momentum space (see cartoon in Fig. 1(a)).
The crossing bands on the mirror plane cannot hybridize since they have opposite mirror eigenvalues, resulting in a stable nodal loop/line on the mirror plane.
The projection of these bulk nodal lines onto the surface fills the inside of the nodal lines and generates the so-called drumhead surface state\cite{Chiu2016,Burkov2011,Zhao2013,Ryu2002,Bian2016,Kim2015,Ma2018}, given its resemblance to the head of an open drum.
The peculiar momentum space structure of these states, flat in both energy and momentum, is critical for the realization of novel phenomena\cite{Tang2014,Kopnin2011,Magda2014,Shapourian2018,Sur2016,Wang2017a,Imada2000,Peotta2015,Huber2010,Wu2007,Tasaki1992,Wang2016,Sun2011,Neupert2011,Pezzini2017,Matusiak2017}, such as topological superconductivity\cite{Shapourian2018,Sur2016,Wang2017a,Kopnin2011,Volovik2015} and magnetism\cite{Magda2014}.
In the case of superconductivity, for example, the momentum extension of the drumhead state (area) is proportional to the pairing strength, and hence $T_\text{c}$\cite{Kopnin2011,Volovik2015}.
Therefore, topological semimetals exhibiting drumhead surface states present a significant expansion of topological materials beyond topological insulators and nodal-point Dirac/Weyl semimetals.\\

Recently, BaAl$_4$ has been reported to have topological semimetallic features with 3D Dirac dispersions and possible nodal lines protected by crystal symmetry\cite{WangMori2021}.
While this can explain several of the observed transport properties, it cannot fully account for the extremely large magnetoresistance, including quantum oscillations\cite{WangMori2021}.
In this paper, we use angle-resolved photoemission spectroscopy (ARPES) to study in detail the electronic structure of BaAl$_4$, focusing on the surface localized bands.
BaAl$_4$ has the body-centered tetragonal structure in the space group of $I4/mmm$ (No. 139) as shown in Fig. 1(b), also known as a prototype parent crystal of a large family of compounds\cite{WangMori2021,Paglione2010,Steglich1979,Stewart1984,Kneidinger2014}.
The bulk and (001) surface projected BZs with high-symmetry points labelled are shown in Fig. 1(c).
The crystal has three non-equivalent mirror-reflection planes $m_{001}$ (green and blue planes), $m_{110}$ (orange plane), and $m_{100}$ (red plane) (see Fig. 1(c)).
The $m_{110}$ plane and $m_{100}$ plane have equivalent mirror planes along the orthogonal directions.
Therefore, a number of Dirac nodal lines can exist on these planes when spin-orbit coupling (SOC) is negligible, leading to the presence of drumhead surface states.\\

Fig. 2 shows the bulk electronic structure of BaAl$_4$.
Panel (a) shows the theoretical bulk nodal lines without SOC.
The energy and momentum dispersions along the high symmetry directions reveal the presence of several nodal points within the valence bands for each mirror plane (see dots in panel (b)).
These nodal points, developing between the highest valence band and the second-highest valence band (Valence Gap (VG)), give rise to a variety of nodal lines in each mirror plane (see colored lines in panel (a)).
The colors used in panels (a) and (b) for the nodal points/lines correspond to the same color scale used to represent the respective mirror plane (Fig. 1(c)).
A detailed analysis of the irreducible representations for each crossing is shown in Supplementary Fig. 1 in the Supplementary Information (SI).
Panels (c) and (d) show the calculated and experimental momentum and energy dispersions along the $\Sigma_1-$Z direction in the $k_z=2{\pi}/c$ plane.
Following the maximum intensity, a Dirac like linear-shaped dispersion can be observed, in agreement with the theoretical calculation in panel (c).
The dispersion can be better extracted by following the peak positions in the momentum distribution curves (MDCs) shown in panel (e) where two peaks disperse linearly throughout the entire energy range and cross at $\sim-0.4$ eV, namely the Dirac point.
The non-gap linear feature is further confirmed in Supplementary Fig. 5 in the SI where the only one side of Dirac dispersion is observed due to the matrix element effect\cite{Chen2012,Damascelli2003}.
Panels (f)-(h) show the same as panels (c)-(e), respectively, but the momentum direction is the $\text{Y}_1-$Z direction in the $k_z=2{\pi}/c$ plane.
Similar to the $\Sigma_1-$Z direction, the Dirac like linear dispersion can also be observed in the $\text{Y}_1-$Z direction and is confirmed by the MDCs spectra, where the two peaks disperse linearly and cross at the ED.
Those Dirac nodes belong to one of the nodal loops in the $k_z=2{\pi}/c$ plane (see the black arrows in panel (a)).
The lack of a gap in the spectra can be due to the absence of hybridization between the two spins and/or weak SOC as in this case, where the bands near $E_\text{F}$ in the VG region are mainly composed of Al $s$ and $p$ orbitals (see details in Supplementary Fig. 2 of the SI).
Once SOC is introduced, the two spins are coupled and allowed to hybridize, resulting in a gap opening at each of the Dirac points/lines.
This is true for each high symmetry directions, unless the $\Gamma-$Z direction where the two bands belong to different representation of the symmetry group, and therefore, their intersection is protected by the crystalline symmetry, $C_{4v}$.
The details of the crystal symmetric information and topological nature with SOC are found in Ref\cite{WangMori2021}.
In the presence of weak SOC, the gap size becomes negligible, and this may be the case for the VG.\\

We now turn our attention to the surface localized electronic states.
Thanks to the matrix element effect, the bulk and surface electronic structures can be characterized selectively (see the detail in Supplementary Fig. 4 in the SI).
Fig. 3 shows the surface electronic structure within the surface BZ projected onto (001) plane.
In addition to the bulk states identified in panel (b) (labelled as B1-B2), three new sets of features are observed in panel (a): sharp linearly dispersive states (labelled as S1-S2), a hole-like dispersion state (labelled as S3), and a weakly dispersive state (labelled as DS state).
Similar features are observed in the surface states calculations shown in panel (b), pointing to their surface origin (see also Method section and Supplementary Fig. 3 in the SI for more details about the calculation).
These multiple surface states, originating from the Dirac nodal lines, indicate their topologically protected nature.
Note that the surface states calculation is sensitive to details of the simulations as reported\cite{Cucchi2020}, and surface effects, such as potential band bending and structural relaxation effects, are not included in the calculation.
These effects might lead to apparent discrepancy between the experimental data (panel (a)) and theory (panel (b)).
Other potential discrepancy could arise from matrix element effects\cite{Chen2012,Damascelli2003}.
These effects can however be minimized by changing photon energy (see Supplementary Fig. 6 in the SI), revealing different features.
The surface origin of these states is further supported by their photon energy dependence (i.e., $k_z$ dependence) as shown in panels (c--f).
Throughout the whole range, negligible dispersions are observed for each of these states, confirming their surface state origin.
Indeed, S1 and S2 states form straight vertical lines as indicated by yellow arrows in panel (c) and blue dashed line in panel (e), indicative of lack of $k_z$ dispersion.
On the other hand, the DS state defines a sheet in the ($k_y$, $k_z$) plane (see dashed yellow rectangle in panel (d)), indicative of a localized state in $k_z$.
This can be directly seen in panel (f), where the energy vs $k_z$ dispersion along the $\text{Z}-\Gamma$ direction shows a localized two-dimensional state.\\

Among all the surface states, of particular interest is the DS state, which is the one appearing at $E=-0.49$ eV.
Indeed, this state emerges out and connects the two Dirac bulk nodal lines, as expected in the case of a drumhead surface state (see also Fig. 3(a)).
Fig. 4 presents its full momentum and energy characterization.
The energy vs momentum dispersion (panel (b)), extracted from the peak position of the energy distribution curves (EDC) (panel (a)), appears weakly dispersing along both the $\bar{\Gamma}-\bar{\text{X}}$ and $\bar{\Gamma}-\bar{\text{M}}$ direction, with an overall bandwidth less than $38$ meV and $23$ meV respectively in the momentum range ($k_1-k_2$).
This gives rise to an almost flat surface state with an effective mass of $m^*\simeq4.0m_e$ and $6.1m_e$ and a peak in the density of state at the energy of the DS state (see the bold line in panel (a)).
In panel (c), we show the momentum extension of the DS state.
The constant energy map ($k_y$ vs $k_x$) at DS state shows that the DS state is localized in momentum along a well-defined filled square-shaped region, centered at the BZ center.
The topology of the DS state is consistent with the confirmed bulk nodal loop (see Fig. 2(a)) and is confined within an area of $\sim$0.16 \AA$^{-2}$, which corresponds to $\sim$8.5 \% of the BZ.
Additionally, the momentum location of the surface states S2-S3 is also visible in this energy window.
In contrast, the main contribution of the S1 state appears near $E_\text{F}$ and is localized along a square-shaped region (see Supplementary Fig. S6 in the SI).
The topology of these states is qualitatively consistent with the theoretical constant energy map shown in panel (d).
All the data reported so far supports the existence of a strongly localized drumhead surface state in BaAl$_4$ and reveals its extended location in momentum space.\\

The localized nature of this state makes it unique with respect to previous studies\cite{Bian2016,Belopolski2019,Lou2018,Muechler2020}, where highly dispersive DS states have been reported.
Indeed the flatness of our DS state drives a large density of states, as shown in Fig. 3--4, enhancing interaction effects significantly.
The large area of the DS state is also promising in view of the enhancing the electrons' interaction\cite{Kopnin2011,Volovik2015}, although the first step would be to engineer the DS state close to the Fermi level.
Finally, it is noteworthy to point out that the observed bulk nodal line might be also responsible for the reported transport anomaly, including the quantum oscillation and extremely large magnetoresistance\cite{WangMori2021}.
A complete quantification of the gap size in BaAl$_4$, and hence the topological classification of BaAl$_4$, requires more detailed calculations and further measurements.
Even a small amount of lattice strain can tune the gap size\cite{Tang2011,Winterfeld2013,Rodin2014,Teshome2018,Owerre2018}, resulting in a new type of Dirac semimetal, where both Dirac point and Dirac nodal lines may coexist.\\

In summary, by combining surface-sensitive ARPES experiments with the theoretical calculations, we have provided direct evidence for the existence of a flat extended surface state in a topological semimetal, BaAl$_4$.
We present that the flat surface state fills inside of the bulk square-shaped nodal line by showing the full momentum space topology of such state.
All the data reported here support that the observed flat band is directly associated with the observed Dirac nodal line, and therefore, the topological drumhead surface states.
These results enable the exploration of such states for the realization of a number of novel correlated phases of matter\cite{Shapourian2018,Sur2016,Wang2017a,Kopnin2011,Volovik2015,Peotta2015,BERGHOLTZ2013,Wang2016,Peotta2015,Sun2011,Neupert2011,Tang2014,Magda2014}.\\

\newpage
\begin{center}{\bf Methods}\end{center}

Single crystals of BaAl$_4$ were synthesized by a high-temperature self-flux method and characterized by X-ray diffraction at room temperature with Cu K$_{\alpha}$ ($\lambda=0.15418$ nm) radiation in a powder diffractometer\cite{WangMori2021}.\\

Electronic structure calculations were performed within the framework of the density functional theory (DFT) with the PAW pseudopotentials, as implemented in the Quantum Espresso package\cite{Giannozzi2009}.
The generalized gradient approximation (GGA) with the Perder-Burke-Ernzerhof parameterization (PBE) was used\cite{Perdew1996}.
A plane wave energy cut-off 40 Ry and 24$\times$24$\times$24 $k$-mesh to sample the BZ were used for the bulk calculations.
Total energies were converged to smaller than $10^{-10}$.
The experimental crystal data ($a=b=4.566$ \AA, $c=11.278$ \AA) were used\cite{Bruzzone1975}.
The calculation of surface electronic structures was carried out with momentum resolved local density of states of a semi-infinite surface by employing a tight-binding (TB) model obtained by using the Wannier90\cite{Mostofi2014} and WannierTools suite of code\cite{Wu2018}.
The quality of the Wannier function based TB model was checked by comparing it with the DFT calculation (see Supplementary Fig. S3 in the SI).
The comparison of our experimental spectra with the theory-calculated surface states shows a good agreement with a 25 \% expansion in the energy dimension of the calculation result.
VESTA package was used for visualization of the crystal structure\cite{Momma2011}.\\

ARPES measurements on single crystalline samples of BaAl$_4$ were performed at the Beamline 4.0.3. end station of the Advanced Light Source in Berkley California.
Samples were cleaved in situ to yield clean (001) surfaces and measured at 20 K in an ultra-high vacuum better than  $3\times10^{-11}$~Torr using the photon energy of 80-128~eV with Scienta R8000 analyzer. The energy resolution was 20-30~meV and the angular resolution was better than $0.2^{\circ}$ for all measurements. According to this photon energy dependence measurements, the inner potential of BaAl$_4$ is estimated to 10.5 eV. \\

\begin{center}{\bf Acknowledgments}\end{center}

This work was primarily funded by the U.S. Department of Energy, Office of Science, Office of Basic Energy Sciences, Materials Sciences and Engineering Division under Contract No. DE-AC02-05-CH11231 (Quantum materials KC2202) and used resources of the Advanced Light Source, a DOE Office of Science User Facility under contract no. DE-AC02-05CH11231.
Research at the University of Maryland was supported by the Gordon and Betty Moore Foundation's EPiQS Initiative through Grant No. GBMF9071, and the Maryland Quantum Materials Center.
TM was supported by JST PRESTO (JPMJPR19L9), and JST CREST (JPMJCR19T3).\\

\begin{center}{\bf Author contributions}\end{center}

A.L., J.P., and R.M. initiated and directed this research project.
R.M. carried out ARPES measurements with the assistance of J.D.D.
R.M. calculated the band structure and analyzed the ARPES data.
A.L. and R.M. wrote the text, with feedback from all authors.
The samples were grown and characterized by K.W.
T.M. provided theoretical insight.

\begin{center}{\bf Competing Interests}\end{center}
The authors declare that they have no competing financial interests.

\begin{center}{\bf Correspondence}\end{center}
Correspondence and requests for materials should be addressed to A.L.~(email: alanzara@lbl.gov).

\begin{center}{\bf Data availability}\end{center}
The data that support the finding of this study are available from the corresponding author upon request.

\bibliographystyle{naturemag}

\newpage
\begin{figure}[H]
  \begin{center}
    \includegraphics[clip,scale=0.35]{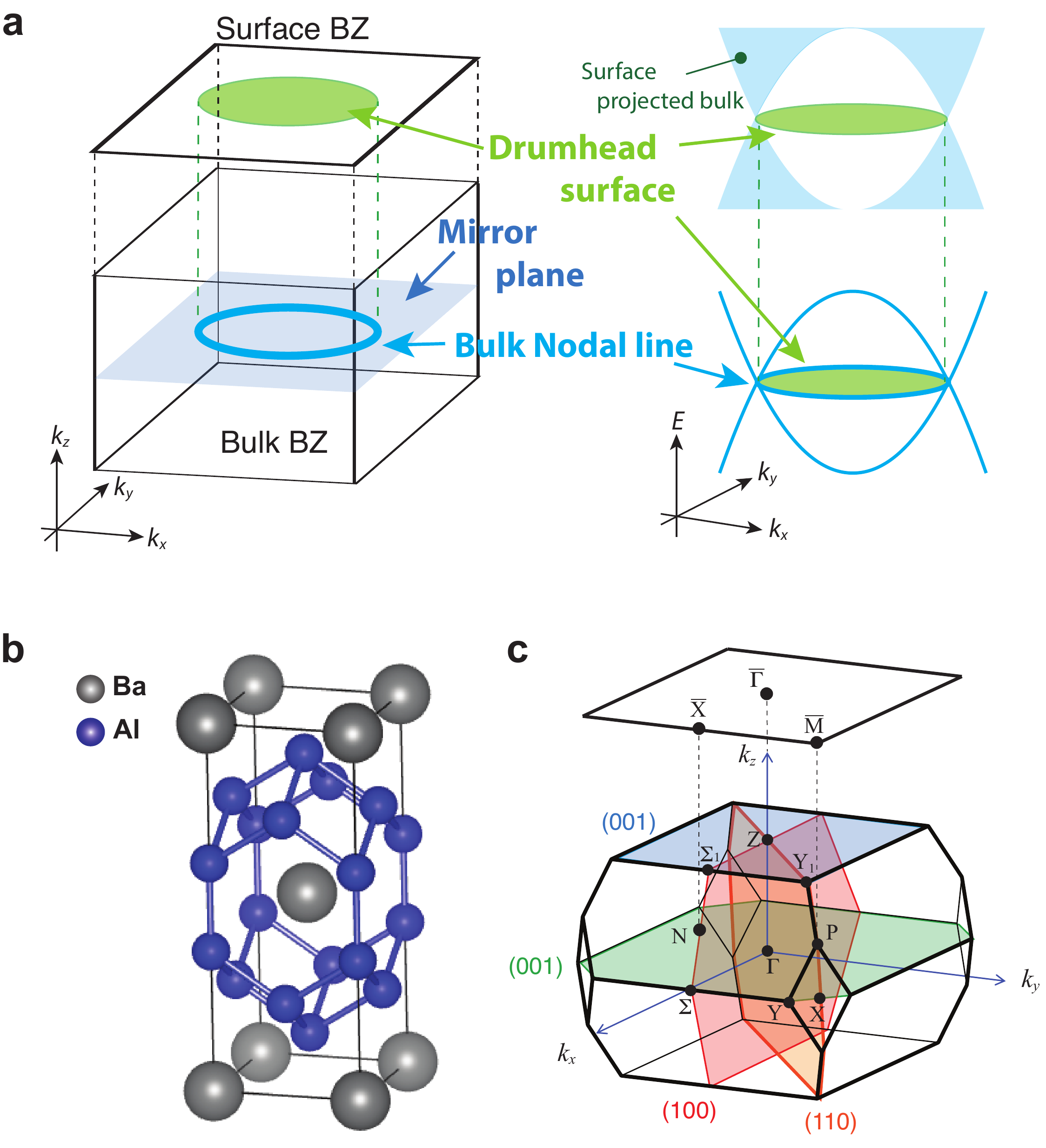}
    \caption{
    	\label{fig:Fig0_comp}
    	\textbf{Topological drumhead surface states and crystal structure of the topological semimetal BaAl$_4$.}
    	\textbf{a,} Schematics of a bulk Dirac nodal line and a drumhead surface state in a topological semimetal in the momentum space (left panels) and in the energy and momentum space (right panels). A gap closing line (light blue solid line) appears in the momentum space, protected by a crystal mirror plane (dark blue plane). Surface states appear inside of the gap closing line, forming the drumhead surface states (light green plane).
    	\textbf{b,} The crystal structure of BaAl$_4$. The gray and the blue spheres represent the Ba and the Al atoms, respectively.
    	\textbf{c,} The bulk Brillouin zone (BZ) and the (001) surface BZ, marked with high-symmetry points. In the bulk structure, the three non-equivalent mirror-reflection planes $m_{001}$ (green plane ($k_z=0$) and blue plane ($k_z=2{\pi}/c$)), $m_{110}$ (orange plane), and $m_{100}$ (red plane) are illustrated.
    	    	}
  \end{center}
\end{figure}
\begin{figure}[H]
  \begin{center}
    \includegraphics[clip,scale=0.42]{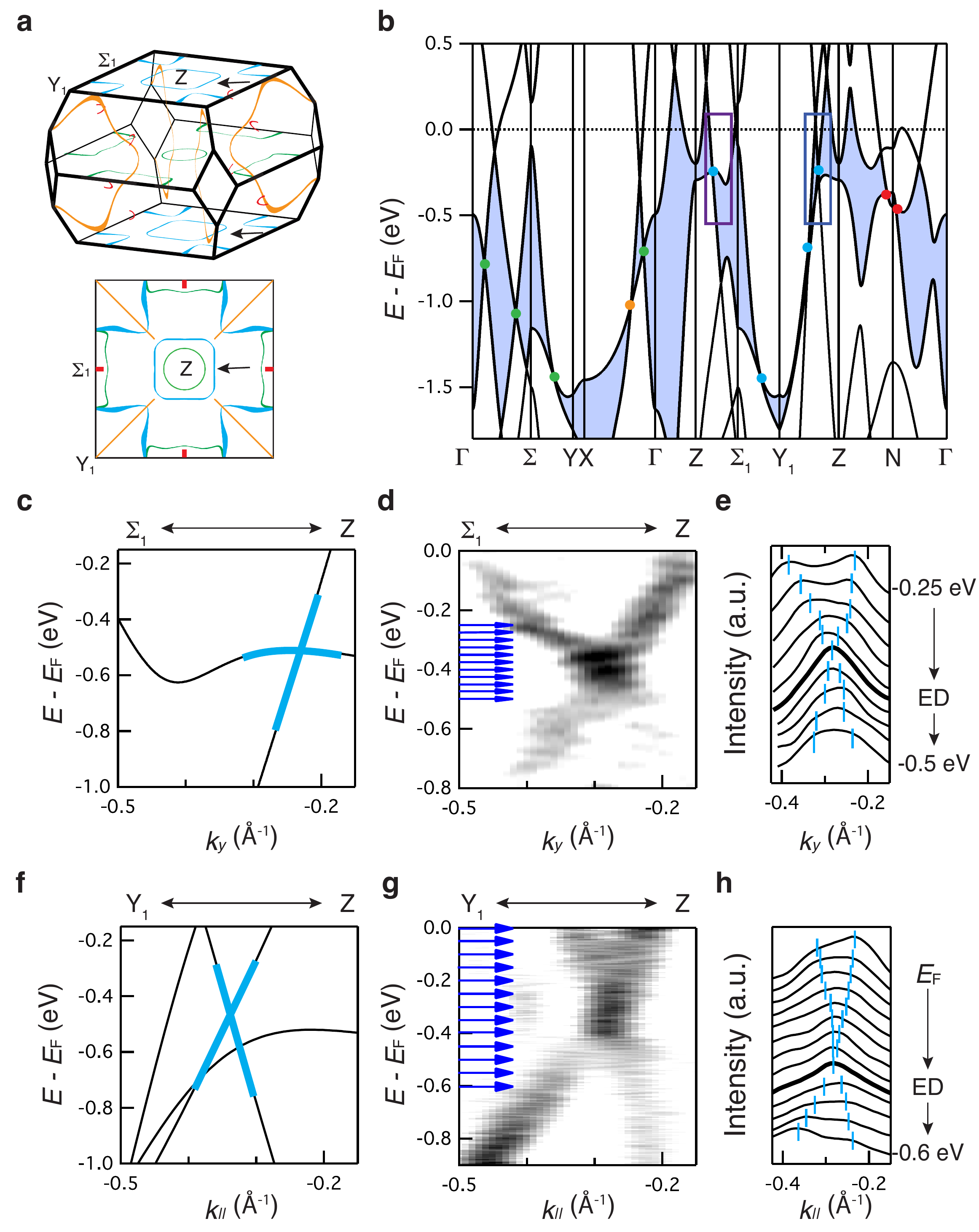}
    \caption{
    	\label{fig:Fig1_comp7_transfer}
    	\textbf{Bulk electronic structure of topological semimetal BaAl$_4$.}
    	\textbf{a,b,} The bulk Dirac nodal lines in bulk BZ (top in \textbf{a}) and on the (001) surface projected BZ (bottom in \textbf{a}) and the calculated bulk electronic structure near $E_\text{F}$ without spin-orbit coupling (SOC) (\textbf{b}). The black arrows in \textbf{a} mark the experimentally identified nodal line in this work. The blue shaded region in \textbf{b} is the energy gap between the highest valence band and the second-highest valence band where the nodal lines shown in this work exist. The blue and purple squares highlight the experimentally observed nodal line in this work. Each color of nodal lines/points in \textbf{a,b} represents the corresponding mirror plane shown in Fig. 1\textbf{c}. 
    	\textbf{c,f,} Zoom in of the calculated electronic structure along $\Sigma_1-$Z (\textbf{c}) and $\text{Y}_1-$Z (\textbf{d}) direction in $k_z=2{\pi}/c$ plane protected by $m_{100}$ (see the purple and blue squares in \textbf{b}). The solid light blue lines in the calculation are guides to the eye for the Dirac band-crossing.
		\textbf{d,g,} The experimental electronic structure (second-derivative in momentum direction) along $\Sigma_1-$Z (\textbf{d}) and $\text{Y}_1-$Z (\textbf{g}) direction, respectively.
        \textbf{e,h,} The momentum distribution curves (MDCs) along the cuts along $\Sigma_1-$Z (see blue arrows from -0.25 eV to -0.5 eV in \textbf{d}) (\textbf{e}) and the cuts along $\text{Y}_1-$Z (see blue arrows from $E_\text{F}$ to -0.6 eV in \textbf{g}) (\textbf{f}), respectively, showing the Dirac dispersion of the nodal line. The light blue marks indicate the peak positions. The bold black lines highlight MDC near the energy of Dirac point (ED).
        }
  \end{center}
\end{figure}
\begin{figure}[H]
  \begin{center}
    \includegraphics[clip,width=\linewidth]{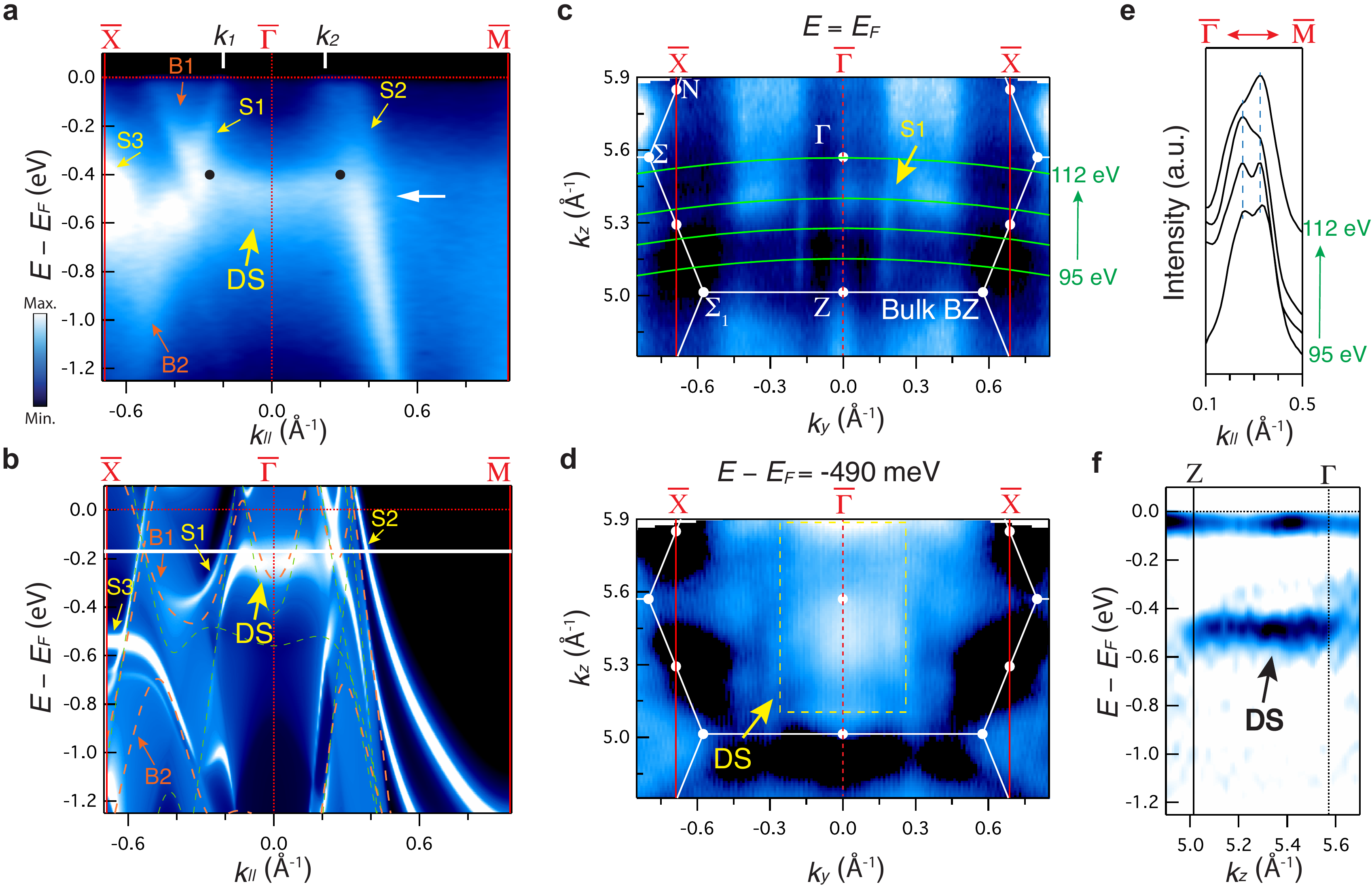}
    \caption{
    	\label{fig:Fig2_AL_12_transfer}
    	\textbf{Surface states of BaAl$_4$.}
	    \textbf{a,} ARPES spectra of energy versus momentum cuts along the high-symmetry directions $\bar{\text{X}}-\bar{\Gamma}-\bar{\text{M}}$ with photon energies 95 eV. The orange and yellow arrows mark the observed bulk (B1-B2) and surface (S1-S3 and DS) states, respectively. The black dots represent the bulk Dirac nodes obtained from the bulk experimental results (Fig. 2).
	    \textbf{b,} Calculated surface band structure along high-symmetry direction $\bar{\text{X}}-\bar{\Gamma}-\bar{\text{M}}$ of BaAl$_4$ (001) surface for a semi-infinite slab. The orange dashed lines represent calculated bulk band structures without SOC for $k_z$ corresponding to the photon energy of 95 eV. The light-green dashed lines are calculated bulk for $k_z=2{\pi}/c$.
       	\textbf{c,} ARPES spectral intensity map in the $k_z-k_y$ plane at the $E=E_\text{F}$. The $k_z$ range covers half of the bulk BZ and corresponds to a photon energy range of 95--112 eV. The green solid lines represent the location of the cuts studied in this work. From bottom to top, each line corresponds to 95, 100, 105, and 112 eV of photon energy, respectively.
       	\textbf{d,} ARPES spectral intensity map in the $k_z-k_y$ plane at the binding energy $E=-490$ meV shown by the white arrow in panel \textbf{a}. The yellow dashed rectangle represents the DS state.
        \textbf{e,} MDCs of spectra along $\bar{\Gamma}-\bar{\text{M}}$ direction at $E=E_\text{F}$ for different photon energies represented in \textbf{c}. The blue dashed lines represent the estimated peak positions.
    	\textbf{f,} Second derivative of energy versus $k_z$ intensity plot.
    	}
  \end{center}
\end{figure}
\begin{figure}[H]
  \begin{center}
    \includegraphics[clip,scale=0.55]{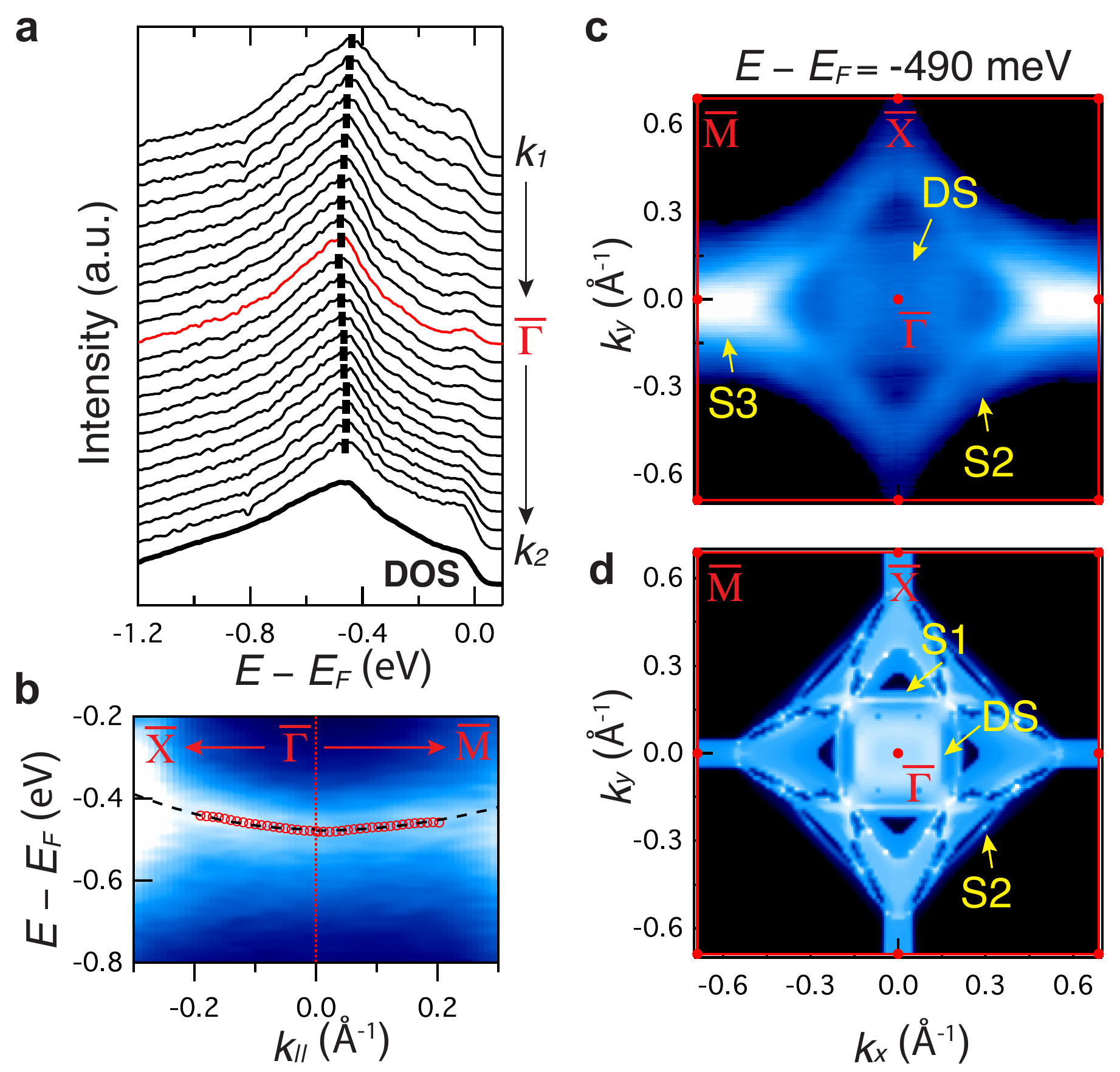}
    \caption{
    	\label{fig:Fig3_AL_comp8_transfer}
        \textbf{Full mapping of the topological drumhead surface state.}
		\textbf{a,} EDCs from $k_1$ to $k_2$ at the position indicated by the marks in Fig. 3\textbf{a}. The bold line at the bottom is the integrated density of state (DOS).
       	\textbf{b,} Extracted dispersions of the drumhead state along $\bar{\text{X}}-\bar{\Gamma}-\bar{\text{M}}$ (the red dots) and the fitting (the black dashed-line).
    	\textbf{c,} Constant energy maps at $E=-0.49$ eV. The surface states (S2-S3) and the drumhead surface (DS) state are marked by the yellow arrows.
       	\textbf{d,} Calculated energy contour at the energy marked by white solid line in Fig. 3\textbf{c}.
    	}
  \end{center}
\end{figure}

\newpage
\newpage
\renewcommand{\figurename}{\textbf{Supplementary Fig.}}
\setcounter{figure}{0}
{\bf Supplementary Note 1: Irreducible representations of crossing points}\\

\begin{figure}[H]
  \begin{center}
    \includegraphics[clip,scale=0.19]{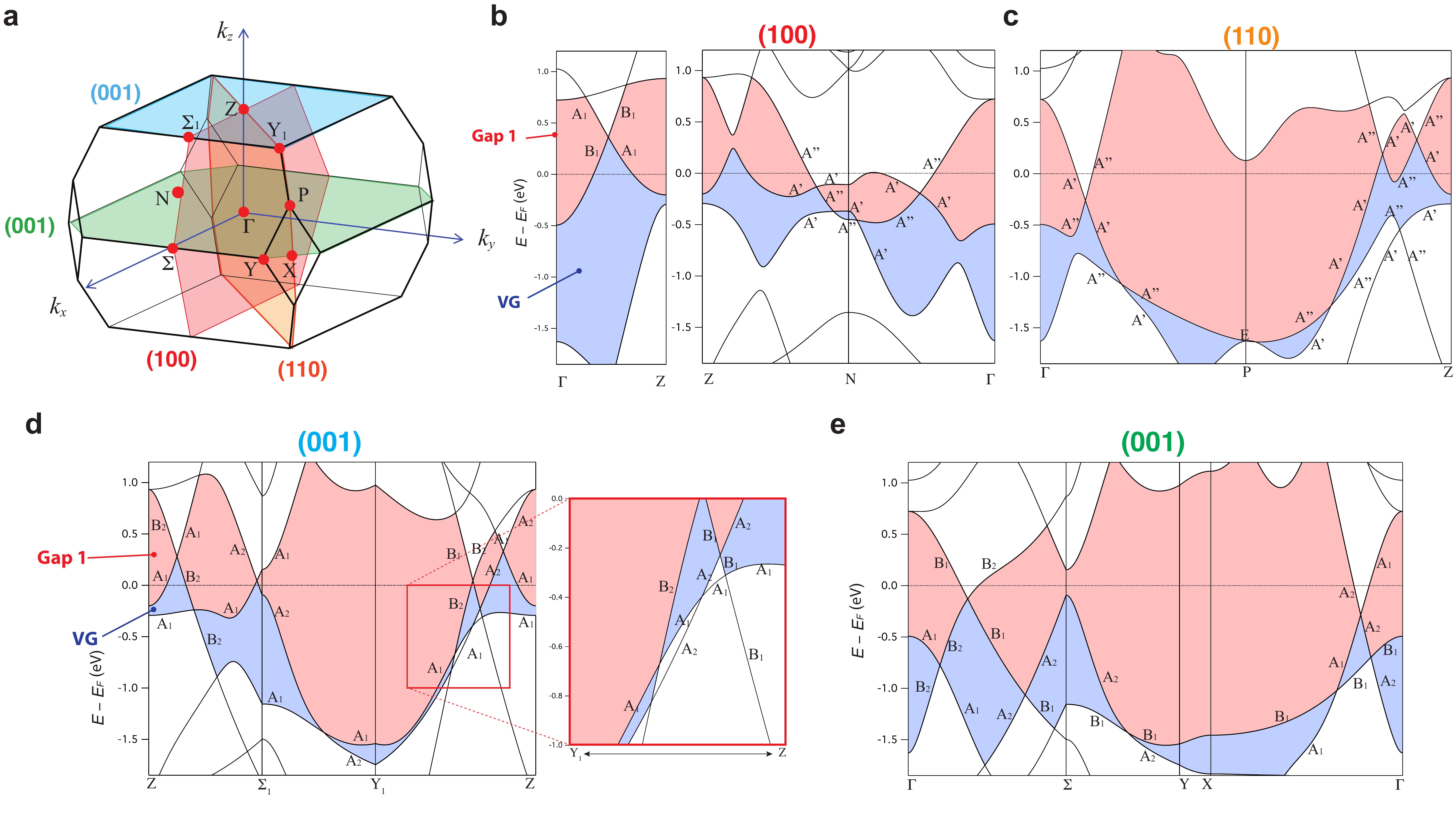}
    \caption{
    	\label{fig:FigS1_comp2}
    	\textbf{Irreducible representations for each crossing band in BaAl$_4$.}
    	\textbf{a} The bulk Brillouin zone (BZ) and the (001) surface BZ, marked with high-symmetry points. In the bulk structure, the three non-equivalent mirror-reflection planes $m_{001}$ (green plane ($k_z=0$) and light-blue plane ($k_z=2{\pi}/c$)), $m_{110}$ (orange plane), and $m_{100}$ (red plane) are illustrated.
    	\textbf{b} The calculated electronic structures with irreducible representations along high symmetry lines on the $m_{100}$ plane; $\Gamma-\text{Z}$ line (left panel) and $\text{Z}-\text{N}-\Gamma$ (right panel). The red shaded region (Gap 1) represents the energy gap between the lowest conduction and highest valence bands and the blue shaded region (Valence Gap (VG)) is the energy gap between the highest valence band and the second-highest valence band. 
    	\textbf{c} The calculated electronic structures with irreducible representations along high symmetry lines on the $m_{110}$ plane. 
    	\textbf{d,e} The calculated electronic structures with irreducible representations along high symmetry lines on the $m_{001}$ plane at $k_z=2{\pi}/c$ (\textbf{d}) and $k_z=2{\pi}/c$ (\textbf{e}) (left panels). Zoom-in of the electronic structure for $\text{Y}-\text{Z}$ (red square in the left panel in \textbf{d}) is shown in the right panel in \textbf{d}. The band crossings on the $m_{100}$ plane, the $m_{110}$ plane, and the $m_{001}$ plane create the band closing line in both Gap 1 (right top panel) and VG (right bottom panel) as represented in Fig. 2(\textbf{b,c}) in the main text.
    	    	}
  \end{center}
\end{figure}

Supplementary Fig. 1 shows the irreducible representations of the band structures for each crossing point.
The crystal structure of BaAl$_4$ has three non-equivalent mirror-reflection planes $m_{001}$ (green and light-blue planes), $m_{110}$ (orange plane), and $m_{100}$ (red plane).
The $m_{110}$ plane and $m_{100}$ plane have equivalent mirror planes along the orthogonal directions.
The crossing bands on these mirror planes cannot hybridize because of the different irreducible representations of the space group where the two crossing bands belong to opposite mirror eigenvalues.
Therefore, these crossing bands form nodal loops/lines on mirror planes.\\

\newpage
{\bf Supplementary Note 2: Orbital contribution near $E_F$}\\
\begin{figure}[H]
  \begin{center}
    \includegraphics[clip,scale=0.4]{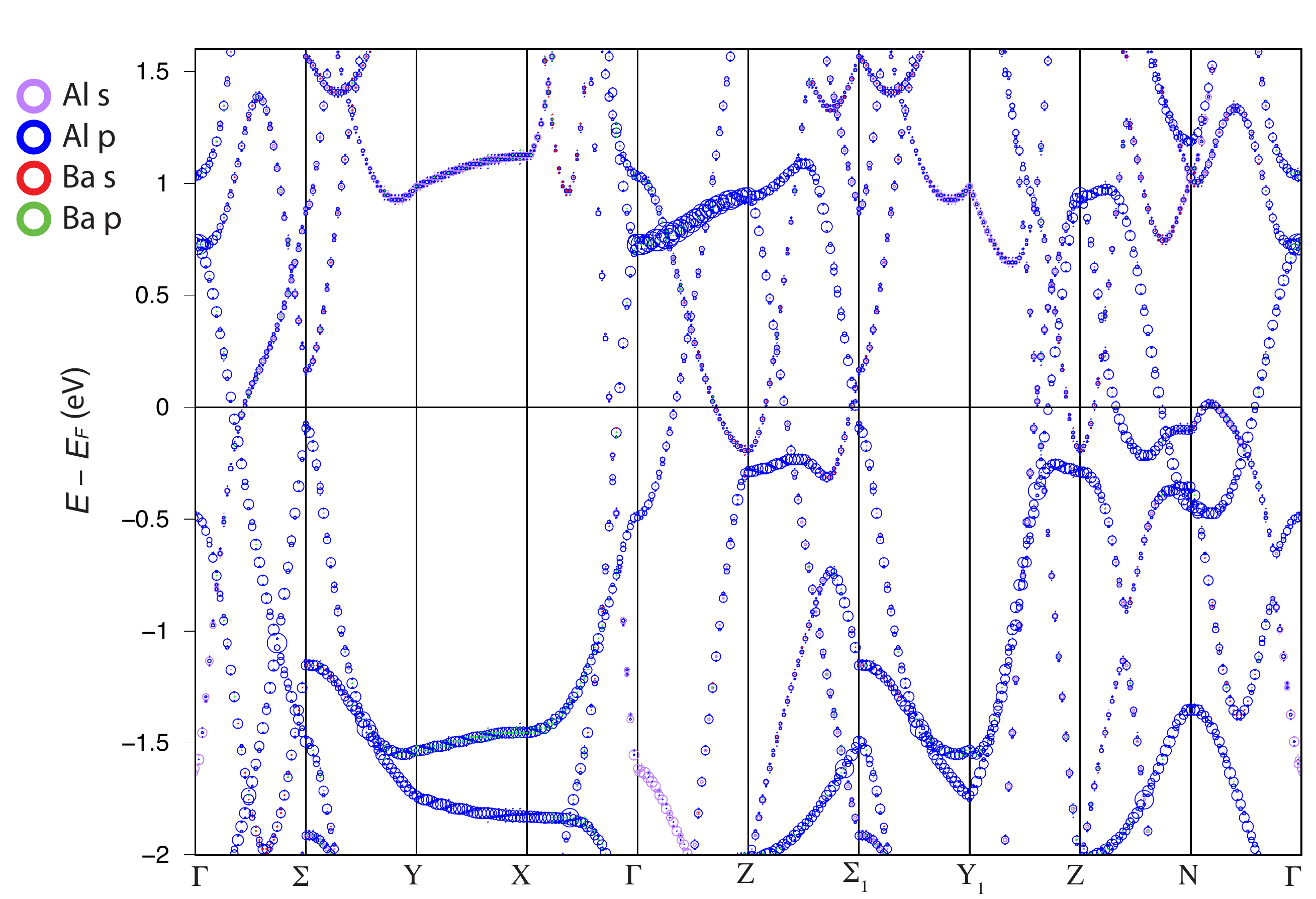}
    \caption{
    	\label{fig:FigS2_comp}
    	\textbf{Calculated electronic structures with orbital weight.}
    	DFT calculated band structure with orbital weight. The size of the dots is proportional to the relative amplitude of the weight of orbital projection onto Al-$s$ (purple), Al-$p$ (blue), Ba-$s$ (red), and Ba-$p$ (green) orbitals. Ba $s$ and $p$ orbitals are represented by red and green. The Fermi level is set to zero.
    	    	}
  \end{center}
\end{figure}

To see the orbital contribution for the band structures near $E_\text{F}$, the orbital projected band dispersions are calculated as shown in Supplementary Fig. 2.
The size of circles represents the relative weight of orbital projection onto Al-$p$, Al-$s$, Ba-$s$, and Ba-$p$ orbitals.
The bands near $E_\text{F}$ are composed mainly of Al-$p$ and Al-$s$ orbital, resulting in small SOC effects.
Once SOC is included in the calculation, hybridization leads to gap opening.
The size of the gap is directly related to the strength of SOC.\\
\newpage
{\bf Supplementary Note 3: Wannier function for tight-binding}\\
\begin{figure}[h]
  \begin{center}
    \includegraphics[clip,scale=0.35]{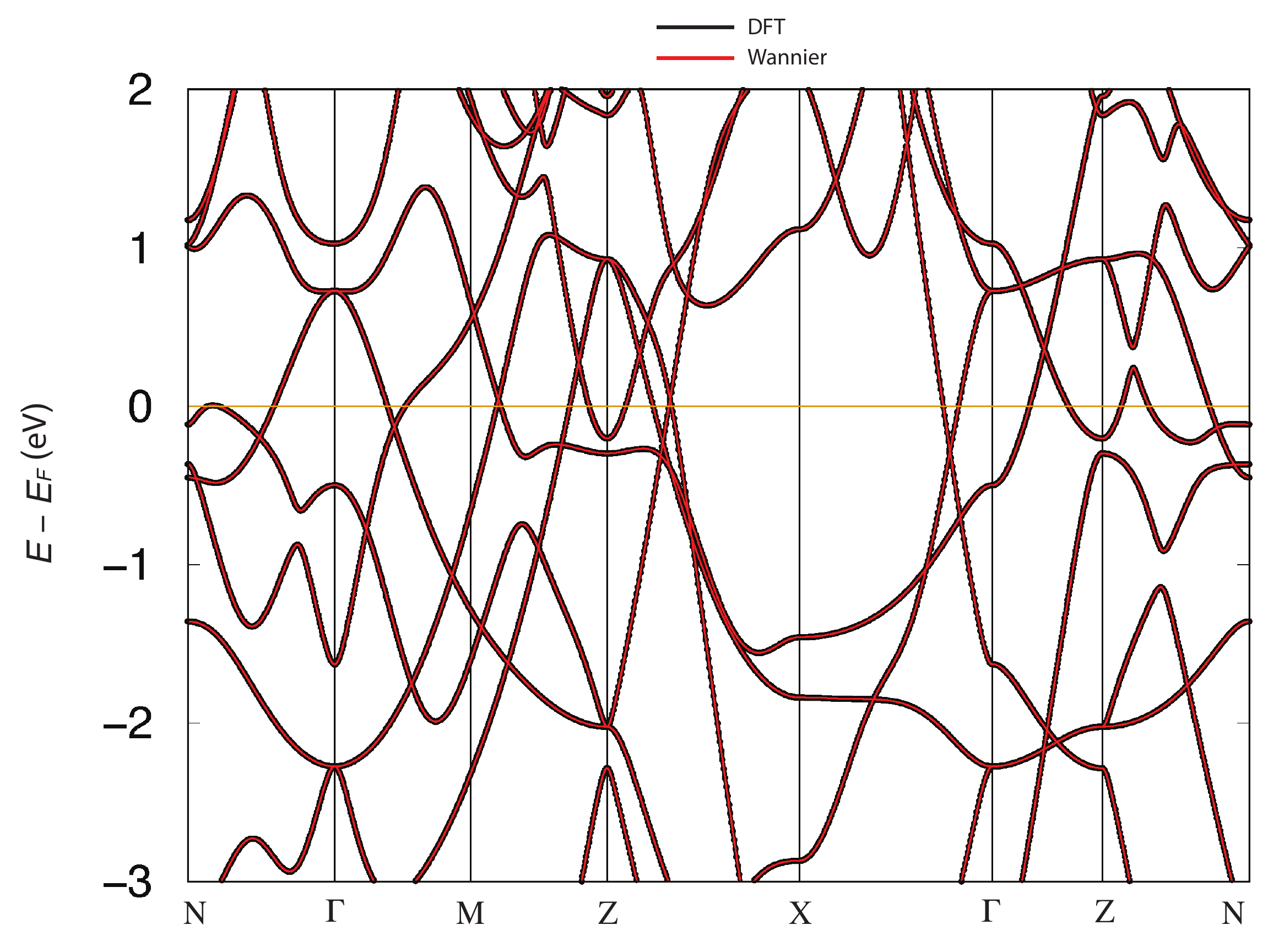}
    \caption{
    	\label{fig:FigS3_comp}
    	\textbf{Comparison Wannier function based tight-binding model with DFT.}
    	Band structure from Wannier function based tight-binding model (red) and DFT (black). Wannier effective tight-binding model reproduces all the features calculated with DFT.
    	    	}
  \end{center}
\end{figure}

To help understand the observed electronic structures in the main text, we create Wannier function based tight-binding model.
Supplementary Fig. 3 shows the band dispersions based on DFT (black lines) and Wannier functions (red lines).
This tight-binding model can well reproduce the energy bands near $E_\text{F}$, which are crucial to the band topology.
This Wannier function-based model is used in the surface electronic structure calculation shown in the main text.\\

{\bf Supplementary Note 4: Bulk and surface electronic structures}\\
\begin{figure}[h]
  \begin{center}
    \includegraphics[clip,scale=0.71]{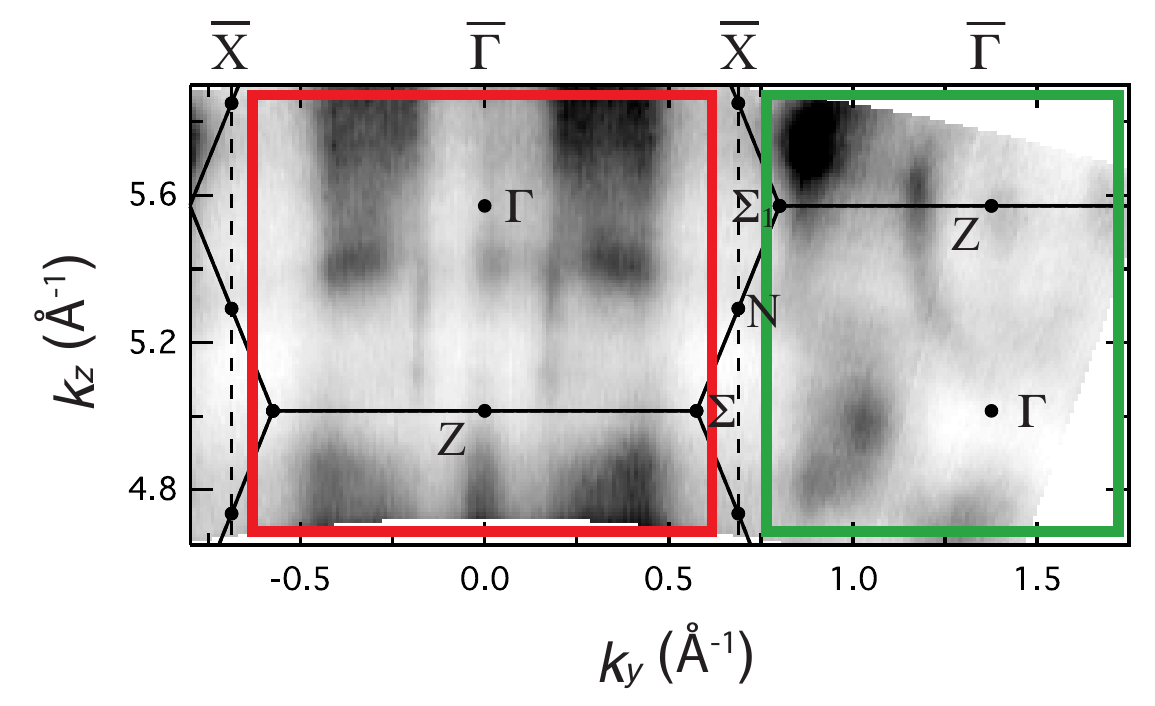}
    \caption{
    	\label{fig:FigS_Fullkz}
    	\textbf{ARPES spectral intensity map in the $k_z-k_y$ plane at the $E=E_\text{F}$ for wide momentum range.}
    	The black solid (dashed) lines represent the bulk (surface) BZ. The red and green square mark the surface and bulk sensitive region in BZ, respectively.
    	    	}
  \end{center}
\end{figure}

Thanks to the matrix element effect, the bulk and surface electronic structures can be characterized selectively.
In Supplementary Fig. 4, we show the out-of-plane Fermi surface mapping taken with the different photon energies.
In the red square region in the panel, negligible dispersions are observed throughout the whole range, confirming their surface state origin.
On the contrary, the dispersive electronic structures are observed in the different BZ (see the green square region), suggesting their bulk state origin.
The observed bulk states here are consistent well with the calculated Fermi surface (see Ref\cite{WangMori2021} for more details about the bulk structure).\\

{\bf Supplementary Note 5: Bulk Dirac nodes}\\
\begin{figure}[h]
  \begin{center}
    \includegraphics[clip,scale=0.5]{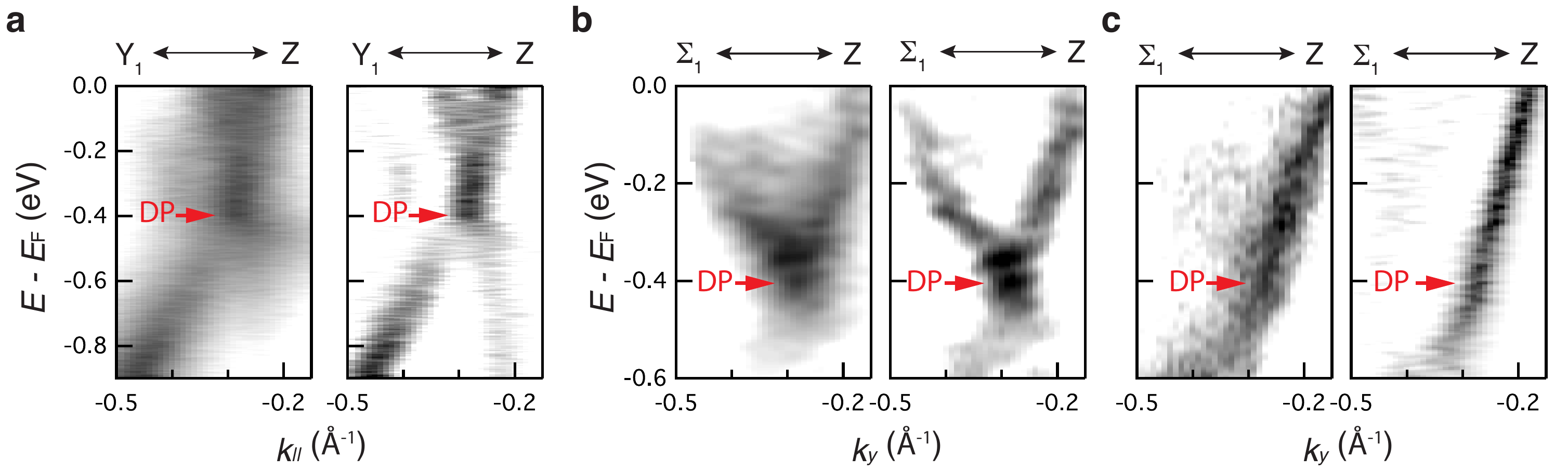}
    \caption{
    	\label{fig:FigS_bulkDirac}
    	\textbf{ARPES spectra of the bulk electronic structure.}
    	\textbf{a} ARPES spectra (left) and its second derivative in momentum (right) along $\text{Y}_1-$Z direction.
    	\textbf{b} ARPES spectra (left) and its second derivative in momentum (right) along $\Sigma_1-$Z direction.
    	\textbf{c} ARPES spectra (left) and its second derivative in momentum (right) along $\Sigma_1-$Z direction, taken in the different BZ from \textbf{b}. The red arrows show the Dirac point (DP).
    	    	}
  \end{center}
\end{figure}

As shown in Fig. 2 in the main text, the bulk Dirac nodes along two high-symmetric lines in $k_z=2{\pi}/c$ plane are observed.
Supplementary Fig. 5(a)-(b) show the raw normalized spectra and their second derivative along $\text{Y}_1-$Z direction. and $\Sigma_1-$Z, respectively.
Panel (c) shows the experimental result along $\Sigma_1-$Z direction (the same direction as panel (b)), but taken in the different BZ.
In this BZ, the only one side of Dirac dispersion is observed due to the matrix element effect, further confirming the non-gap linear dispersion.\\

{\bf Supplementary Note 6: Photon energy dependence of bulk and surface states}\\

\begin{figure}[h]
  \begin{center}
    \includegraphics[clip,scale=0.33]{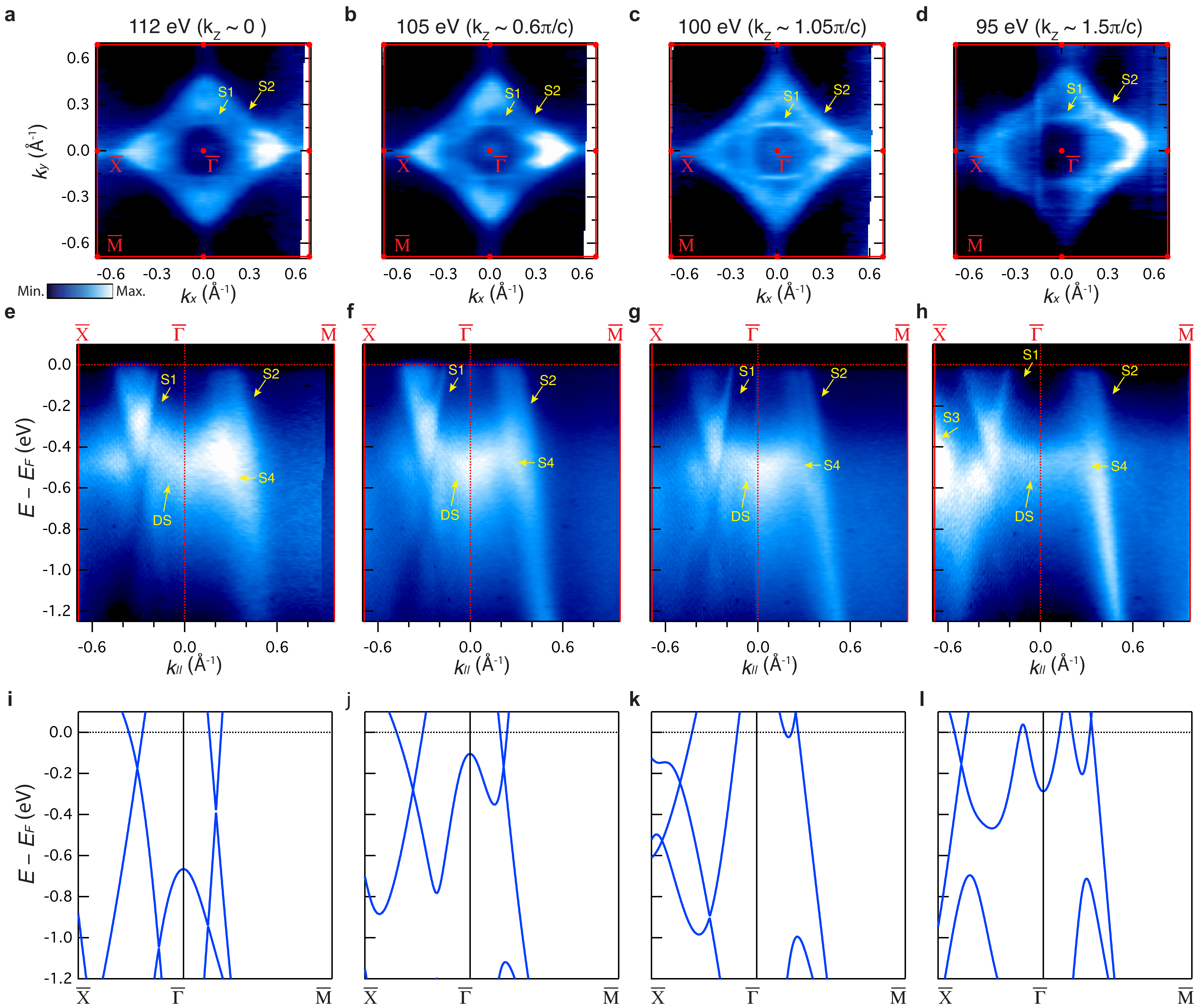}
    \caption{
    	\label{fig:FigS4_comp}
    	\textbf{ARPES spectra with different photon energies and corresponding calculated bulk bands.}
    	\textbf{a--d} Fermi surface mappings with different photon energies 112 (\textbf{a}), 105 (\textbf{b}), 100 (\textbf{c}), and 95 (\textbf{d}) eV.
    	\textbf{e--h} ARPES spectra of energy versus momentum cuts along the high-symmetry directions with photon energies 112 (\textbf{a}), 105 (\textbf{b}), 100 (\textbf{c}), and 95 (\textbf{d}) eV.
    	\textbf{i--l} Calculated bulk band structures without SOC. Each structure is calculated for $k_z$ corresponding to each photon energy; $k_z\sim0$ ($h\nu=112$ eV in (\textbf{i})), $k_z~\sim-0.6\pi/c$ ($h\nu=105$ eV in (\textbf{j})), $k_z~\sim-1.05\pi/c$ ($h\nu=100$ eV in (\textbf{k})), and $k_z~\sim-1.5\pi/c$ ($h\nu=95$ eV in (\textbf{l})).
    	    	}
  \end{center}
\end{figure}

Photon energy dependence is a powerful way to extract the $k_z$ dependence of electronic structures.
The $k_z$ dependence can be obtained by changing photon energy using the free-electron final state approximation:
\begin{equation}
k_z = \frac{1}{\hbar}\sqrt{2m_e(E_i+h\nu-\Phi)\cos\theta+V_0}
\end{equation}
where $m_e$ is the free electron mass, $E_i$ is the energy of the initial state, $h\nu$ is the photon energy, $\Phi$ is the work function, $\theta$ is the emission angle of photoelectrons, and $V_0$ is the inner potential.
Electronic structures taken with the different photon energies resolve the surface nature of the observed spectra; surface states are two-dimensional states and hence showing no $k_z$ dependence, while bulk states show $k_z$ dispersive three-dimensional feature.

Supplementary Fig. 6 shows the experimental spectra (panels (a--h)) and the calculated bulk structures for several different photon energies, corresponding to different $k_z$ values in panels (i--l); $k_z\sim0$ ($h\nu=112$ eV in (\textbf{i})), $k_z~\sim0.6\pi/c$ ($h\nu=105$ eV in (\textbf{j})), $k_z~\sim1.05\pi/c$ ($h\nu=100$ eV in (\textbf{k})), and $k_z~\sim1.5\pi/c$ ($h\nu=95$ eV in (\textbf{l})).
The Fermi surface mappings (panels (a)-(d)) confirm that the topology of the observed surface states (S1-S2) near $E_\text{F}$ is consistent with the theoretical constant energy map shown in Fig. 4(d) in the main text.
By comparing the experimental spectra (panels (e)-(h)) and the bulk calculations (panels (i)-(l)), the S1--S4 states and the DS state are clearly distinguished from the bulk electronic features, validating their surface origins.
Note that the matrix element effects play a different role for different photon energy, leading to the different appearance of these states.
Indeed, S4 appears in panels (e)-(g) and the intensity is suppressed in panel (h), and S3 appears in panel (h).	
\end{document}